\documentclass[sigconf]{acmart}

\AtBeginDocument{%
  }

\setcopyright{acmcopyright}
\copyrightyear{2023}
\acmYear{2023}
\acmDOI{XXXXXXX.XXXXXXX}

\usepackage{bm}
\usepackage{amsmath,scalerel}

\DeclareMathOperator*{\concat}{\scalerel*{\Vert}{\sum}}
\begin{document}


\title{AQ-GT: a Temporally Aligned and Quantized GRU-Transformer for Co-Speech Gesture Synthesis}

\author{Hendric Voß}
\email{hvoss@techfak.uni-bielefeld.de}
\affiliation{%
  \institution{Social Cognitive Systems Group}
  \institution{Bielefeld University}
  \streetaddress{Universitätsstraße 25}
  \country{Germany}
}

\author{Stefan Kopp}
\email{skopp@techfak.uni-bielefeld.de}
\affiliation{%
  \institution{Social Cognitive Systems Group}
  \institution{Bielefeld University}
  \streetaddress{Universitätsstraße 25}
  \country{Germany}
}

\renewcommand{\shortauthors}{Voss and Kopp}

\begin{abstract}
The generation of realistic and contextually relevant co-speech gestures is a challenging yet increasingly important task in the creation of multimodal artificial agents. Prior methods focused on learning a direct correspondence between co-speech gesture representations and produced motions, which created seemingly natural but often unconvincing gestures during human assessment. 
We present an approach to pre-train partial gesture sequences using a generative adversarial network with a quantization pipeline. The resulting codebook vectors serve as both input and output in our framework, forming the basis for the generation and reconstruction of gestures. By learning the mapping of a latent space representation as opposed to directly mapping it to a vector representation, this framework facilitates the generation of highly realistic and expressive gestures that closely replicate human movement and behavior, while simultaneously avoiding artifacts in the generation process. We evaluate our approach by comparing it with established methods for generating co-speech gestures as well as with existing datasets of human behavior. We also perform an ablation study to assess our findings. The results show that our approach outperforms the current state of the art by a clear margin and is partially indistinguishable from human gesturing. We make our data pipeline and the generation framework publicly available.
\end{abstract}




\maketitle

\section{Introduction}
Effective communication between humans encompasses various modalities, such as speech, facial expressions, and bodily gestures. The ability to comprehend and generate these multimodal signals enables us to engage in meaningful and nuanced conversation and is commonly employed in everyday interactions \cite{cassell1999speech,wagner2014gesture}. Consequently, researchers have devoted considerable efforts to accurately process, interpret, and generate these communicative cues to facilitate natural and seamless human-machine interaction \cite{koppCommonFrameworkMultimodal2006a}.

Generating non-verbal cues, such as body language and hand gestures, to accompany spoken language poses a particularly challenging task for embodied agents. In recent years, machine learning approaches have been developed to predict gestures based on a given linguistic input. Such co-speech-driven gesture synthesis frameworks achieve impressive outcomes in generating realistic gestural motion that aligns with a given spoken utterance or text but still face difficulties in generating movements that meet the communicative effectiveness and richness of human gesture  \cite{nyatsangaComprehensiveReviewDataDriven2023b}.

In this paper, we introduce a framework that aims to learn a discrete latent space of gesture sequences. The proposed AQ-GT approach utilizes a combination of Gated Recurrent Units (GRU) and Transformers to learn an intermediate representation of vectors, which, through a temporally aligning network architecture, constructs new novel co-speech gestures. We thereby make several contributions to the field of gesture synthesis. 
Firstly, we propose a novel data processing pipeline that enables the tracking of full 3D body joints for arbitrary monocular videos. This pipeline enables the automatic acquisition of precise gesture information, facilitating the learning of natural gestures.
Secondly, we introduce a novel approach to co-speech gesture synthesis. This approach leverages a discrete latent space of gestures and combines the strengths of gated recurrent units (GRUs) and Transformers to generate naturalistic gestures that are closely aligned with accompanying speech.
Thirdly, we present a thorough evaluation of our framework, using both objective and subjective measures, demonstrating that it outperforms current state-of-the-art approaches and generates gestures that are partially indistinguishable from human gesturing.
Finally, we present an ablation study that isolates and analyzes the contribution of single components of the proposed model to the overall output quality. A video with examples of generated co-speech gestures is made available online\footnote{\href{https://vimeo.com/823756031}{https://vimeo.com/823756031}}.

\section{Related Work}

The generation of co-speech gestures has been a long-standing goal for socially interactive multimodal agents such as embodied conversational agents, 3D avatars, or socially assistive robots. Early approaches mainly relied on rule-based models that were crafted by hand, based on empirical or theoretical insights. One such example is the Behaviour Markup Language \cite{koppCommonFrameworkMultimodal2006a} that allowed for the generation of multimodal behavior by specifying function-to-form mappings in an XML format. Similarly, the BEAT toolkit \cite{cassellBEATBehaviorExpression2004} employed expandable rule sets, that were derived from linguistic and contextual analyses of human conversational behavior. Later methods applied dynamic learning approaches that mainly used kernel-based probability distribution models or stochastic Markov models to learn motion patterns from motion capture sequences or high-level structures from temporal segmentation \cite{brandStyleMachines2000a,galataLearningVariableLengthMarkov2001}. Due to their high degree of customization and low computational requirements, such techniques have also been used in commercially available products such as the Pepper and Nao robots \cite{pandeyMassProducedSociableHumanoid2018}.

In recent years, the development of deep machine learning has significantly impacted the field of multimodal behavior processing and virtual agent behavior generation \cite{changchunliuEmpiricalStudyMachine2005, kimPororobotDeepLearning}. As a result, data-driven approaches have become the primary method for generating complex behaviors in various human-agent and human-human interaction tasks \cite{yuInteractiveRobotLearning2019a, tapusPerceivingPersonTheir2019a}.

During the early stages of deep learning, the main objective was to train virtual agents to demonstrate natural behaviors and create visually appealing co-speech gestures, without the need for time-consuming hand-crafted techniques. The main focus was on using a single input modality. Initial attempts such as a study conducted by \citet{fanTTSSynthesisBidirectional2014a} utilized a Bidirectional Long Short-Term Memory (LSTM) model to generate new sequences of co-speech gesture solely from text input and a given initial gesture input. Similarly, \citet{ferstlInvestigatingUseRecurrent2018a} employed a recurrent neural network with an encoder-decoder architecture, trained on prosodic speech features to generate short sequences of motion, while  \citet{ginosarLearningIndividualStyles2019a} proposed a model to generate gestures from speech audio data along with an initial pose.

More recent work includes multiple input modalities, each exhibiting different dimensionalities and structures, and has prompted the development of more complex data-driven methodologies. In their investigation of co-speech gestures, \citet{holdenPhasefunctionedNeuralNetworks2017a} employed feed-forward networks, training them with a cyclical function to generate gestures. In contrast, \citet{henterMoGlowProbabilisticControllable2020} applied a distinct deep-learning technique, implementing a glow network with invertible 1x1 convolutions for co-speech gesture generation. \citet{lingCharacterControllersUsing2020a} adopted a related approach, utilizing a variational autoencoder combined with deep reinforcement learning for goal-oriented control of co-speech gestures.

Current research highlights the potential of general adversarial networks (GANs) to produce highly realistic results, with the capacity to merge multiple modalities cohesively, thus enhancing the representation of human communication. \citet{ahujaStyleTransferCoSpeech2020} devised a method for learning style embeddings, enabling the generation of unique gesture styles for individual speakers. This approach enables a diverse range of gesture variations, which could be dynamically adjusted during the generation process.
To integrate audio, text, and speaker identity within a GAN framework, \citet{yoonSpeechGestureGeneration2020} designed a technique that effectively generates hand gestures closely resembling those observed in natural human communication. This method produces gestures that reflect the semantic meaning and rhythm of spoken words, thus simulating authentic human interaction in virtual environments.
Likewise, \citet{aoRhythmicGesticulatorRhythmAware2022a} recently introduced an approach to gesture synthesis that accounts for both rhythmic and semantic aspects. Specifically, they integrated a rhythm-based segmentation pipeline with neural embeddings of speech and motion, drawing on insights from linguistic theory.

\section{Data}
Several datasets are currently available for training co-speech gesture generation algorithms. However, they often have certain limitations \cite{nyatsangaComprehensiveReviewDataDriven2023b}. For instance, many datasets contain only a small number of gesture sequences, while others provide solely 2D tracking of body joints or do not incorporate tracked hand information such as finger movements.
To the best of our knowledge, only two datasets provide 3D body joints and are sufficiently large for our purposes: the BEAT dataset \cite{liuBEATLargeScaleSemantic2022a} and the Talking with Hands dataset \cite{leeTalkingHands162019,yoon2022genea}. Unfortunately, both of these datasets cover only a limited number of speakers, which may restrict a model's ability to generalize across speaker styles. Therefore, we decided to collect and build our own dataset, called the BiGe dataset.

\subsection{Data Acquisition}
In order to successfully train co-speech gesture generation algorithms, the given data must meet certain requirements. In particular, large amounts of noise or significant variability in the data can have an adverse effect on the generated gestures. Since we aimed to capture full-body gestures and especially hand gestures of the presenter, we needed to ensure that the entire person is visible. Additionally, the person had to face the camera to avoid obscuring their gestures. Given the challenge of eliminating camera movements, we also required a video source with stationary cameras. To meet these criteria, we opted to use the same data source as \citet{yoonRobotsLearnSocial2019a}, namely the official TED channel on Youtube \cite{ted_youtube}. To diversify our dataset with multi-language presentations, we also selected the TEDx Talks Youtube channel which features international presentations in different languages \cite{youtubeTEDxTalks}. In total, we collected 6945 videos with the highest possible video and audio format, as well as automatically and manually annotated subtitles. Next, we filtered out any videos that did not have available subtitles, featured a resolution lower than 1280x720 pixels, or lacked subtitles with word-for-word timing information. After the filtering process, we were left with 4327 videos for our dataset, with a total length of 1021 hours. 

\subsection{Data Processing}

\begin{table}[ht]
  \centering
  \caption{\textmd{\small{Overview of the BiGe dataset and the Shots of Interest (SoT)}}}
  \resizebox{\linewidth}{!}{
  \begin{tabular}{|l|l|}
    \toprule
    Number of videos & 4327 \\\hline
    Total length of videos & 1021 h \\\hline
    Videos with SoT & 2756 \\\hline
    Total length of SoT & 260.6 h \\\hline
    Average length of SoT & 17 s  \\\hline
    Number of SoT & 54.360 (20 per video on average) \\
    \bottomrule
  \end{tabular} \label{tbl:SoT}
  }
  \label{tab:table_groundtruth}
\end{table}

To extract body joint keypoint information from the collected videos, a modified data processing pipeline was created, based on the work of \citet{yoonRobotsLearnSocial2019a}.
In order to track 59 3D full-body joints for each individual in the videos, we employed a combination of the AlphaPose library \cite{alphapose}, the VideoPose3D model \cite{pavllo20193d}, and the MediaPipe hands model \cite{zhang2020mediapipe}.
The pipeline first utilized a YOLOv3 model \cite{redmon2018yolov3}, through the AlphaPose library, to detect all persons in each frame of the video. Then, the 2D position of each joint was estimated through the implementation of a FastPose model \cite{zhang2019fast}. The tracked joints, excluding the hands, were subsequently transformed into 3D coordinates by employing the VideoPose3D model \cite{pavllo20193d}. Finally, the MediaPipe Hands model was used to determine the 3D positions of the hands by elevating the previously tracked 2D hand joints to 3D coordinates. 
After tracking, we cut the videos into multiple clips based on detected camera cuts. We then determined Shots of Interest (SoT) for the entire dataset, which are any clips that fulfill the removal conditions established by \citet{yoonRobotsLearnSocial2019a}. As the tracking pipeline provided us with a confidence score for each joint, we used this score to further refine our selection of SoT by removing any SoT where the median value falls below a threshold $\tau$, which we set to 0.05. Table \ref{tbl:SoT} gives an overview of the characteristics of the resulting BiGe dataset: It comprises 2756 videos, out of which we extracted 54.360 SoTs with an average length of 17 seconds. To create our dataset, we split these data into 2156 training videos, 300 validation, and 300 test videos.

\section{AQ-GT Architecture}
\begin{figure*}[bth]
  \centering
  \includegraphics[width=15cm]{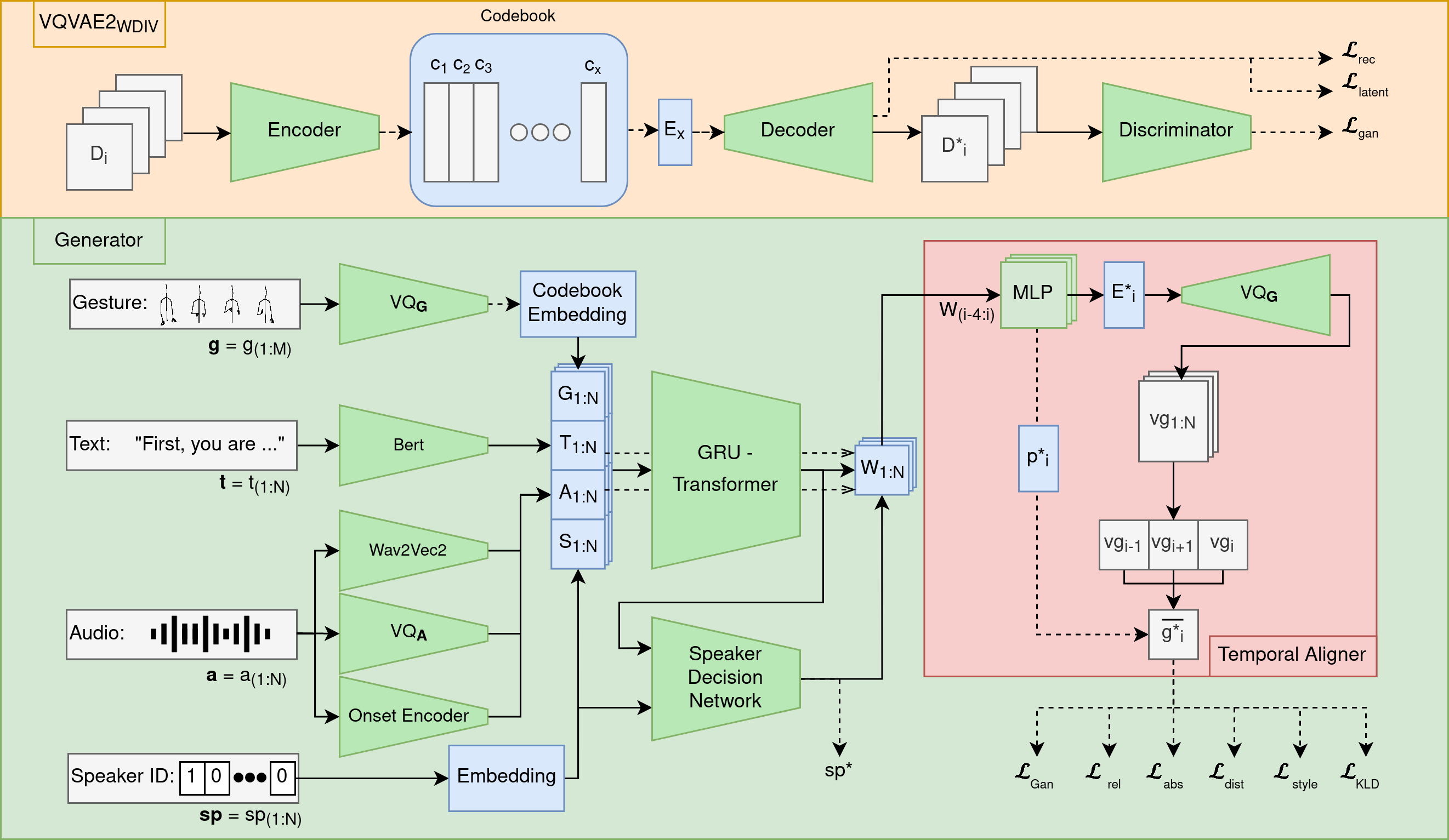}
  \caption{Overview of the AQ-GT model. Top: the VQVAE2 network with the added GAN discriminator. Bottom: The generator network with input modalities, pre-processing, GRU-Transformer and Temporal Aligner network (in red).}
  \label{fig:model_overview}
\end{figure*}

The overall structure of the proposed model architecture is shown in Figure \ref{fig:model_overview}. The model consists of two parts: The first one (colored in yellow) is a quantized variational autoencoder that yields a structured embedding space representation of sequential data, both for the raw audio as well as the gestural data. The second one (colored in green) is the model for autoregressively generating gesture frames from audio, text, and speaker ID input. This part comprises a preprocessing and integration of the input data, a GRU-Transformer, a network for speaker prediction, and a temporal aligner. We emphasize the use of the temporal aligner as well as the quantized latent space for the GRU-transformer in the name of the model, "AQ-GT". Each of the components and their training is described in the following.

\subsection{Quantization in Discrete Latent Space} \label{discrete_latent}
Drawing inspiration from the work of \citet{esserTamingTransformersHighResolution2021}, we utilize a combination of a Vector Quantized Variational Autoencoder (VQ-VAE) \cite{van2017neural} and a Generative Adversarial Network (GAN) \cite{goodfellow2020generative,liu2019acceleration}. We employ the VQ-VAE-2 model \cite{razavi2019generating} in conjunction with a Wasserstein Generative Adversarial Network with Divergence penalty (WGAN-div)  \cite{wu2018wasserstein} to learn a discrete latent space of concise gesture sequences.
The original VQ-VAE model \cite{van2017neural} consists of two main components, an encoder and a decoder, both of which employ a shared "codebook". The encoder maps observations onto a sequence of discrete latent variables, while the decoder reconstructs the original observations using these discrete variables. To achieve this, the encoder uses a non-linear mapping from the input space, $x$, to a vector $E(x)$, which is then quantized based on its distance to the prototype vectors $c_i, i \in 1 . . . C$ in the shared codebook. 
The VQ-VAE-2 model introduced by \citet{razavi2019generating} can be seen as an extension of the VQ-VAE, which incorporates a hierarchical structure. In this model, a codebook is learned for each hierarchical level, which we denote as $C_{top}$, $C_{bottom}$, and the combination of both codebooks as $C$. The loss function used for training the VQ-VAE-2 is given by:
\begin{equation}\label{eq:loss_vqvae_normal}
\mathcal{L}_{vq}(\mathbf{x}, D(\mathbf{z}))  = ||\mathbf{x} - D(\mathbf{z})||_2^2 + ||sg[E(\mathbf{x})] - \mathbf{z}||_2^2 + \alpha ||sg[\mathbf{z}] - E(\mathbf{x})||^2_2
\end{equation}
where $E$ represents the Encoder, $D$ represents the Decoder, $sg$ refers to the stop-gradient, and $\alpha$ is a hyperparameter that controls the impact of the regularisation term. The first term in Equation \ref{eq:loss_vqvae_normal} measures the reconstruction error between the input and the output of the decoder, while the second and third terms encourage the model to use the codes from the codebook and to minimize the distance between the encoder output and the selected codes, respectively.

One notable issue encountered in quantized models is the occurrence of codebook collapse, wherein a substantial portion of the codebook is disregarded, and only a limited subset is utilized. Although the VQ-VAE-2 model enhances the stability of the codebook by incorporating hierarchical learning, this problem may still manifest. Recent work by \citet{esserTamingTransformersHighResolution2021} has demonstrated that the combination of a VQ-VAE with a GAN results in a perceptually more diverse codebook and more stable training. In light of these findings, we add a discriminator for the adversarial training of the VQ-VAE-2 model, which processes the decoded output of the VQ-VAE-2 and outputs an unbounded scalar value. Leveraging the insights from the research of \citet{wu2018wasserstein}, we utilize a Wasserstein divergence objective for the training which enhances the fidelity and stability of the output, as compared to Wasserstein loss with gradient penalty \cite{gulrajani2017improved}. We call this combination of the VQ-VAE-2 and the WGAN-div the VQ-VAE-2\textsubscript{wdiv} model.

The loss function for our discriminator is given by:
\begin{equation}\label{eq:loss_vqvae_wgan_div_dis}
\mathcal{L}_{D_{wdiv}}(\mathbf{x}, D(\mathbf{z}))  = Dis(\mathbf{x}) - Dis(D(\mathbf{z})) + \delta \|\nabla_{\mathbf{\hat{x}}} Dis(\mathbf{\hat{x}})\|^p
\end{equation}
where $Dis$ is the discriminator and $\delta$ is a hyperparameter that determines the strength of the divergence penalty. The first part of the loss $Dis(\mathbf{x}) - Dis(D(\mathbf{z}))$ evaluates the difference between the real sample $\mathbf{x}$ and the output of our VQ-VAE-2 $D(\mathbf{z})$. The second term $\delta \|\nabla_{\mathbf{\hat{x}}} Dis(\mathbf{\hat{x}})\|^p$ is the divergence penalty, which encourages the generated sample $D(\mathbf{z})$ to be close to the real data distribution.

The generator loss function is simply given by
\begin{equation}\label{eq:loss_vqvae_wgan_div_gen}
\mathcal{L}_{G_{wdiv}} = Dis(D(\mathbf{z}))
\end{equation}
which minimizes the result of the discriminator output. Therefore, our complete loss function for the generator of the VQ-VAE-2 model is the combination of the $\mathcal{L}_{vq}$ and the $\mathcal{L}_{G_{wdiv}}$ loss functions: 
\begin{equation}\label{eq:loss_vqvae_full}
\mathcal{L} = \beta\mathcal{L}_{vq} + \gamma\mathcal{L}_{G_{wdiv}}
\end{equation}
We additionally introduce the hyperparameter $\beta$ and $\gamma$ to adjust the contribution of both loss functions.
 
\subsection{Input Modalities} \label{modalities}
In our proposed framework, we leverage a combination of three input modalities (initial or previous gesture position, text, and audio), along with a speaker ID. Each of these modalities is pre-processed in a specific way. To facilitate the flexibility of our framework, we allow for arbitrary lengths of initial information and generated gesture sequences. To this end, we introduce two parameters: $N$ represents the number of frames in the initial information, which we set to four; $M$ denotes the number of frames generated by the gesture sequence, which we set to 30. To ensure consistency in our pre-processing pipeline, we sample gestures at a rate of 15 frames per second, resulting in a 2-second sequence of newly generated gestures.

For the pre-processing of previous gesture frames $g$, we train a VQ-VAE-2\textsubscript{wdiv} model on the entire gesture training set, with a learning rate of $2e-5$ and a batch size of $256$. Henceforth, we will call this model $VQ_G$. For the introduced hyperparameters we perform a hyperparameter Bayesian search and, based on the results, set $\alpha$, $\beta$, and $\gamma$ to $0.25$, $1$, and $1$, respectively. After training for 200 epochs we restore the weights with the lowest validation loss and freeze the model weights. Instead of using the absolute position in 3D space, we use the relative position of the parent bone as our input and output vectors. Finally, we construct an embedding space for each of the 512 possible codebook vectors and, based on the indices of the codebooks, return the selected embeddings.

The text pre-processing uses the pre-trained BERT model \cite{devlinBERTPretrainingDeep2019}, which converts the given text $tex$ to tokens and extracts a high-level representation of it. For our framework, we return the representation vector with a fixed length of up to 300 tokens.

The audio processing pipeline employs three distinct models to effectively process and extract meaningful features from the raw audio data $a$. These models include the Wav2Vec2 framework \cite{baevskiWav2vecFrameworkSelfSupervised2020}, our custom pre-trained VQ-VAE-2\textsubscript{wdiv} model, and an Onset Encoder model \cite{liangSEEGSemanticEnergized2022}.
The Wav2Vec2 framework is a semi-supervised model trained on speech input, which we use to process the raw audio data and obtain the last hidden state of the model. However, as the focus of the Wav2Vec2 framework is primarily on reconstructing speech from given audio data, it may result in the loss of crucial prosodic features during training. To address this, we additionally train our own VQ-VAE-2\textsubscript{wdiv} model on 0.25-second segments of downsampled 16kHz raw audio data. We collect 1500 hours of audio data from randomly selected YouTube videos, which we split into 1200 hours of training data and 150 hours of validation and test data. We use the same learning rate, batch size, and hyperparameter configuration as for the $VQ_G$. As the dataset is far larger than our gesture dataset, we train the model for 100 epochs and afterward freeze the weights with the best validation loss. We refer to this model as the VQ\textsubscript{A} model. During the training of the entire framework, we process the audio with this model and return the vector $E_x$.

Recent studies have highlighted the significance of sound onset in co-speech gesture generation \cite{liangSEEGSemanticEnergized2022,aoRhythmicGesticulatorRhythmAware2022a}. By leveraging this information, the model can synchronize the generated gestures with the audio input more effectively, resulting in a closer adherence to the rhythm of the speech. To this end, we adopted the Onset Encoder model proposed by \citet{liangSEEGSemanticEnergized2022}. This model first uses a filter to detect the onset of the audio and then decouples the speech input into semantically relevant and semantically irrelevant cues. Unlike the other two models, the weights of the Onset Encoder model are not frozen and updated during the framework training. 

To account for different speaker styles, we use the speaker identity $sp$ and create an embedding space for each possible speaker in the dataset. After passing the embeddings through a multilayer perceptron (MLP), we employ the "Reparameterization Trick" as proposed by \citet{kingma2013auto} to establish a multivariate normal distribution within the latent space. This technique can be leveraged to generate new, unseen gesture styles, as demonstrated in prior works  \citep{ahujaStyleTransferCoSpeech2020,ghorbani2022exemplar}.
In addition to the embedding space, we define a "Speaker Decision Network" consisting of an MLP network, which predicts the current speaker identity $sp^*$ given the input vector $W$ and returns a corresponding embedding layer based on the result. We train this network using a Softmax contrastive loss as described in Section \ref{training_proc}.

\subsection{GRU-Transformer}

\begin{figure}[b]
  \centering
  \includegraphics[width=\linewidth]{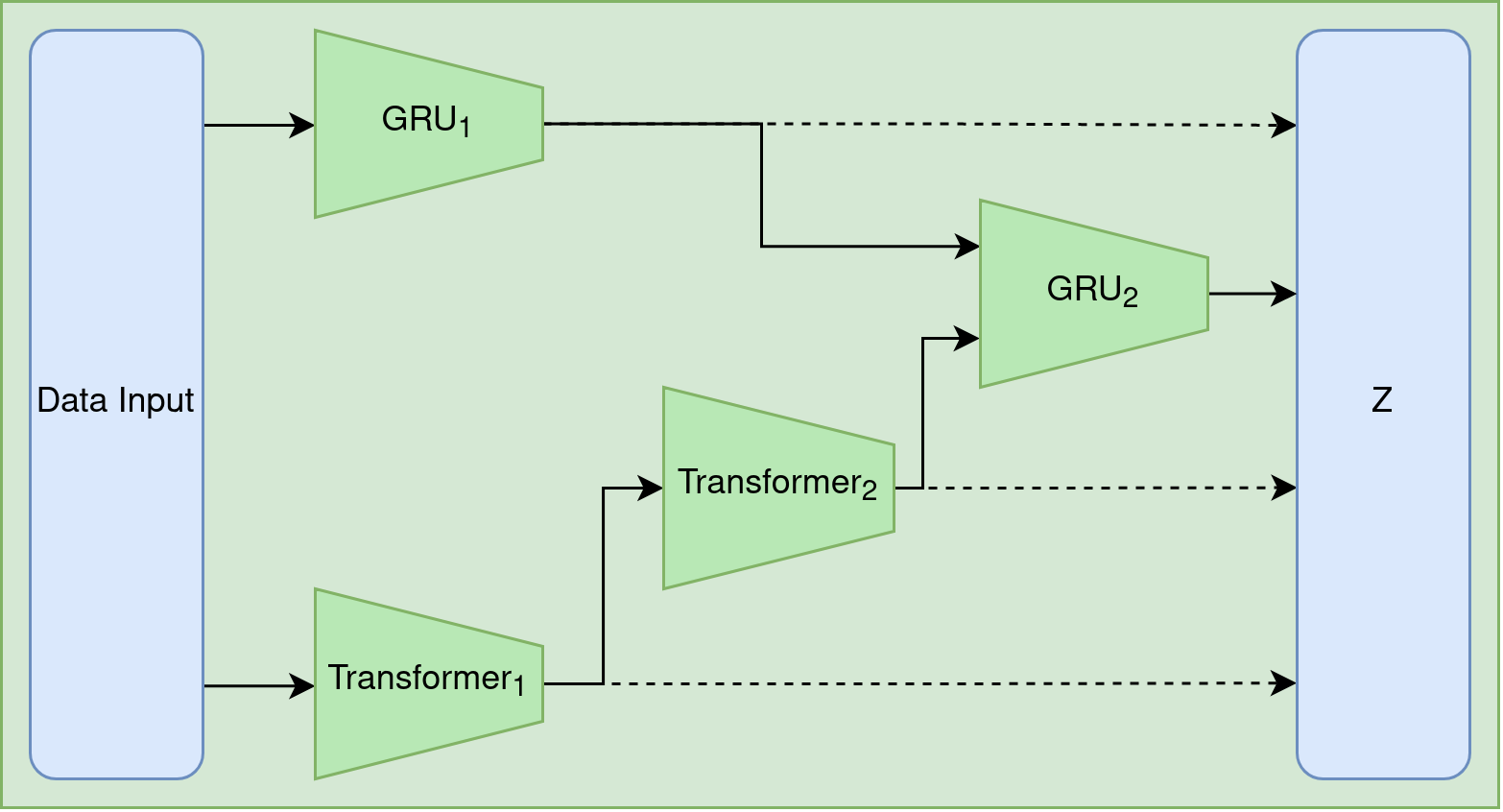}
  \caption{Architecture of GRU-Transformer component. The input and output data are colored in blue; model layers are colored green; dotted lines represent skip connections.}
  \label{fig:gesture_gru_transformer}
\end{figure}

The data-based generation of co-speech gestures requires the modeling of temporal dependencies in sequential gesture data. A popular choice for this are Gated Recurrent Units (GRUs) \cite{cho2014learning} because of their high versatility and ease of use \cite{yoonSpeechGestureGeneration2020,liangSEEGSemanticEnergized2022,liuLearningHierarchicalCrossModal2022a,aoRhythmicGesticulatorRhythmAware2022a}. The particular effectiveness of GRUs lies in their ability to capture non-linear dependencies, which is particularly important in modeling sequential gesture data that often exhibit complex temporal dynamics. However, the performance of GRUs is often hindered by challenges such as slow convergence rates, limited learning efficiency, and vanishing gradient problems \cite{wang2019ogru}. At the same time, Transformer architectures \cite{vaswani2017attention} have been introduced to learn long-term dependencies in structured input data. Despite their potential, transformers are not commonly utilized in co-speech gesture generation for their high computational demands and limited capabilities of processing complex, non-linear gesture data \cite{ahuja2020no,bhattacharya2021text2gestures,zhuang2022text}. 

We propose to fuse these two approaches to comprehensively and robustly model the complex temporal (sequential) structure of co-speech gestures. In particular, we use the Transformer to capture the broader global dependencies within the input sequence, while the GRU enables the learning of non-linear, localized dependencies.
As shown in Figure \ref{fig:gesture_gru_transformer}, we propose a GRU-Transformer model that processes the data input both with a GRU and a transformer pipeline. The transformer blocks are comprised of a multi-headed self-attention block with temporal information pre-processing, followed by a multilayer perceptron (MLP) to reduce the dimension of the output. As the maximum number of heads has to be divisible by the input, we set the number of heads to $n_{heads} \leftarrow max(\{n \in {1, 2, 3, ..., 12} : n \mod n_i = 0 \})$, where $n_i$ is equal to the input dimension. 
The GRU\textsubscript{1} block is comprised of a single GRU layer followed by a layer normalization. For the GRU\textsubscript{2} block we use four stacked GRU layers.
Using skip connections, the intermediate output of all blocks is added to the output vector.

\subsection{Temporal Aligner}
One of the challenging tasks in generating co-speech gestures is creating temporal synchronicity between the relevant verbal and gestural events. Although both transformers and GRUs are able to learn temporal information from the input, in practice these layers still struggle to create convincing gestures that are free of sudden unwanted gesture artifacts \cite{liuLearningHierarchicalCrossModal2022a,fan2020addressing}. Therefore, we add a temporal aligner network, which learns the temporal dependencies of the previously generated frames and aligns the reconstructed sequences to remove sudden noise in the generated gestures. The structure of our Temporal Aligner is visualized in Figure \ref{fig:model_overview} (red square). To construct the aligner, we reuse the VQ\textsubscript{g}, described in Section \ref{modalities}. After receiving the combined vector $W$, we apply a sliding window to iterate over the preceding three frames and concatenate them with the current frame and the result of the previous iteration. We then pass the combined vector into an MLP and save the result for the next iteration. Formally, we compute the temporal vector
\begin{equation}
    f_c(\mathbf{W}_t) = MLP(f_c(\mathbf{W}_{t-1}) +\!\!\!+ \concat_{i=0}^{-3} \mathbf{W}_{t-i}   )
\end{equation}
From this, we reconstruct the vector $E_x$ of the network $VQ_G$, denoted as $E^*_t$, for each time step $t$. Using this vector $E^*_t$ we reconstruct the next gesture frame $vg_t$ with the $VQ_G$ network. As the $VQ_G$ is trained on a sequence of four frames, we compose the gesture vector $g^*_t$ out of the 
third element of $vg_{t-1}$, the second element of $vg_{t}$, and the first element of $vg_{t+1}$. Formally, we perform:
\begin{equation} \label{eq:1}
g^*_t = \frac{1}{3}\sum_{i=0}^2 vg_{t-1+i,2-i}
\end{equation}
Once the vector $g^*_t$ has been constructed, we average it with the vector $p^*_t$, which is computed using an MLP given the output of the temporal vector $f_c(\mathbf{W}_t)$. This ensures that the network can correct noise remaining in the reconstruction of the vector $g^*_t$.

\section{Training the Generator Model} \label{training_proc}
The generator model is trained as illustrated in Figure~\ref{fig:model_overview}. In the model, the separately pre-processed input data is combined into an input vector for the GRU-Transformer. Using its output we predict the current speaker identity using a Speaker Decision Network. We combine the output of the GRU-Transformer, the Speaker Decision Network, and the input text and audio into a vector $W$. Using this vector, we synthesize the gesture $G^*$ through a Temporal Aligner. 

We train this model with the same WGAN-div approach as described in Section \ref{discrete_latent}.
For the training of the framework, we define multiple loss functions to guide the different parts of the training. When calculating a distance between two sets of vectors, we use the Huber Loss \cite{huberRobustEstimationLocation1964}, which we denote as $HL$. 

First, we define the absolute and relative reconstructive loss. For this, we compare the Huber distance given by the output of our generator model compared to the ground truth, as well as the absolute Huber distance given by reconstructing the skeleton structure of the pose data, both for the generated gestures and the ground truth. Formally, we define the reconstructive loss as the average Huber distance between the skeleton of the generated poses and the ground truth skeleton:
\begin{align}
    \mathcal{L}_{abs} &= \frac{1}{N} \sum^N_{i=1} LH(Rc(g_i),Rc(g^*_i))
    \mathcal{L}_{rel} &= \frac{1}{N} \sum^N_{i=1} LH(g_i,g^*_i)
\end{align}
where $Rc$ is the reconstruction function for the absolute joint position, $N$ is the sample size and $g^*_i$ is the gesture vector from the generator model. 

To ensure correct gestures involving both hands, we define a loss that measures the pairwise distance between all bones in the left and right arms and compares it to the ground truth. Formally, we define the distance loss as:
\begin{align}
\mathcal{L}{dist} = \frac{1}{N}\sum_{i=1}^N\sum_{j=1}^{K}HL(\mathbf{g_i}_{l,j}-\mathbf{g_i}_{r,j},\mathbf{g^*_i}_{l,j}+\mathbf{g^*_i}_{r,j})
\end{align}
where $K$ is the number of arm joints.

We also add a speaker contrastive loss, employing a combination of the speaker style loss proposed by \citet{yoonSpeechGestureGeneration2020} and our speaker decision softmax loss. Our training approach involves generating gestures based on input $T$ and the original speaker identity $\bm{sp}{id(1)}$, while maximizing the distance for a random speaker identity $\bm{sp}{id(2)}$ with the same input $T$. Formally, we define:
\begin{align}
    f_{st} &= \mathrm{CE}(\mathbf{G}(\bm{T},\bm{sp}_{id(1)}),\mathbf{G}(\bm{T},\bm{sp}_{id(2)})) \\
    f_{sp} &= \mathrm{HL}(\mathbf{SPD}(\bm{T},\bm{sp}_{id(1)}),\mathbf{SPD}(\bm{T},\bm{sp}_{id(2)})) \\
    \mathcal{L}_{\mathrm{style}} &= -\mathbb{E}\left[\min \left(\frac{f_{st} + f_{sp}}{\|\bm{sp}_{id(1)} - \bm{sp}_{id(2)}\|_1}, \epsilon\right)\right]
\end{align}
where $SPD(T,sp)$ is the softmax output of the Speaker Decision Network and $CE$ is the Cross-Entropy Loss. To make it easier to sample different styles, we use the Kullback-Leibler (KL) divergence loss ($\mathcal{L}_{KLD}$) between the feature space of the speaker identity and $N(0,I)$ to assume a Gaussian distribution for the style embedding.

Finally, we need to account for the position and temporal dependencies of joint vectors in natural gesturing. To that end, we include the first to sixth derivatives of the joint position vector with respect to time in a loss function. Formally, we define:
\begin{align} \label{eq:dop}
f_{dop}(g) &= \frac{1}{6} \sum_{i = 1} ^6 \frac{d^ig}{dt^i}\\
\mathcal{L}_{DoP} &= \frac{1}{N} \sum^N_{i=1} LH(f_{dop}(g_i),f_{dop}(g^*_i))
\end{align}

The overall loss function combines the components as follows:
\begin{align}
\mathcal{L}{total} &= \pi_1 \mathcal{L}_{gan} + \pi_2 (\mathcal{L}_{rel}+\mathcal{L}_{abs}) \\
&+ \pi_3 \mathcal{L}{dist} + \pi_4 \mathcal{L}_{\mathrm{style}} + \pi_5 \mathcal{L}_{KLD} + \pi_6 \mathcal{L}_{DoP}
\end{align}
with $\pi_1,\pi_2,\pi_3,\pi_4,\pi_5,\pi_6$ being hyperparameter to adjust the loss strengths. We performed a hyperparameter search for all parameters of our framework. Based on the results, we use a batch size of $190$, $\pi_1$ of $2$, $\pi_2$, $\pi_3$, $\pi_4$, and $\pi_6$ of $20$ and $\pi_5$ of $0.004$. We use an unlimited number of epochs combined with a stop criteria to finish the training and restore the best-performing weights if there is no decrease in the validation loss after 50 epochs. Subsequently, we train our model for 368 epochs and restore the weights from epoch 318. 
We adopt a step-wise decreasing learning rate schedule to optimize the performance of our model. Specifically, we initialize the learning rate to $1e-4$. Every 20 epochs we multiply the learning rate by a factor of $0.75$, to a minimum of $1e-5$. To reduce the possibility of overfitting, we use dropout with a step-wise increment of 0.05 every 25 epochs, up to a maximum of 0.3, between every layer and for the input of the network.

\begin{table}[bt]
    \caption{Comparison of the proposed model with three state-of-the art models for the BiGe dataset. For FGD and MAJE scores lower is better; for Diversity distance higher is better.}
  \centering
  \begin{tabular}{lccccc}
    \toprule
    Methods & FGD $\downarrow$ & Diversity $\uparrow$ & MAJE $\downarrow$\\
    \midrule
    Ground Truth & 0.00 & 42.128 & 0.00\\
    \midrule
     Trimodal \cite{yoonSpeechGestureGeneration2020} & 2.023 & 38.980 & 0.0119\\
     SEEG \cite{liangSEEGSemanticEnergized2022} & 2.018  & 38.660 & 0.0121\\
     HA2G \cite{liuLearningHierarchicalCrossModal2022a}   & 1.346  & 41.332 & 0.0096\\
     \midrule 
     \textbf{Ours (AQ-GT)} & \textbf{0.3977} & \textbf{42.885} & \textbf{0.0078}\\
    \bottomrule
  \end{tabular}
  \label{tbl:res}
\end{table}

\section{Evaluation}
To evaluate the proposed framework, we compare the generated gestures with what other state-of-the-art frameworks produce, both using available objective measures and with a user study to investigate subjectively perceived differences between the generated output.

\begin{table}[b]
    \caption{Comparison of the proposed model with state-of-the art systems on the TED dataset. For FGD lower is better; for the Diversity distance higher is better (SEEG and Rythmic Gesticulator  do not report a Diversity distance).}
  \centering
  \begin{tabular}{lcccc}
    \toprule
    Methods & FGD $\downarrow$ & Diversity $\uparrow$ \\
    \midrule
    Ground Truth & 0.00 & 110.821 \\
    \midrule
     Joint Embedding \cite{ahujaLanguage2PoseNaturalLanguage2019}   & 22.083  & 91.223\\
     Speech2Gesture \cite{ginosarLearningIndividualStyles2019a}  & 19.254  & 98.095\\
     Attention Seq2Seq \cite{yoonRobotsLearnSocial2019a} & 18.154 & 92.176\\
     SEEG \cite{liangSEEGSemanticEnergized2022}   & 3.751 & -\\
     Trimodal \cite{yoonSpeechGestureGeneration2020}   & 3.729 & 102.539\\
     HA2G \cite{liuLearningHierarchicalCrossModal2022a}   & 3.072 & \textbf{108.086}\\
     Rythmic Gesticulator \cite{aoRhythmicGesticulatorRhythmAware2022a}  & 2.04 & -\\
     \midrule 
     \textbf{Ours (AQ-GT)} & \textbf{1.612} & 104.76\\
    \bottomrule
  \end{tabular}
  \label{tbl:ted_res}
\end{table}

\begin{figure*}[t]
  \centering
  \includegraphics[width=12cm]{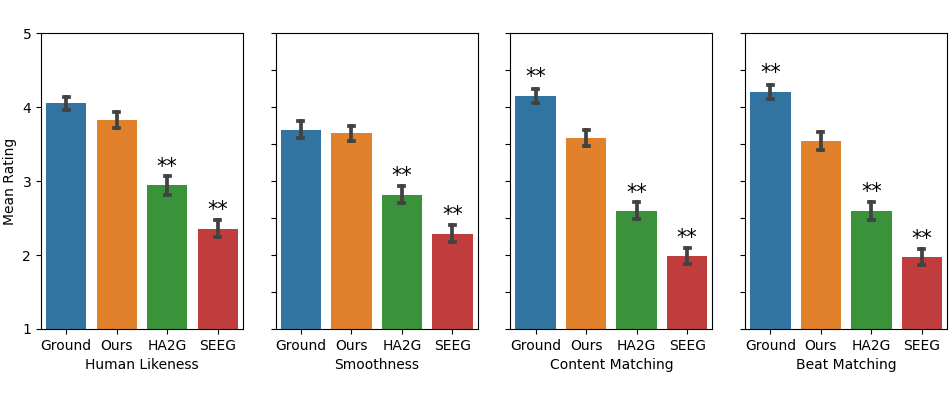}
  \caption{The results of the subjective evaluation study. The ground truth is marked in blue. Our framework is marked in yellow. The HA2G framework is marked in green. The SEEG framework is marked in red. Asterisks indicate significant effects (\text{*} : p < 0.05, \text{*}\text{*} : p < 0.005).}
  \label{fig:study_result}
\end{figure*}

\subsection{Objective Evaluation}
To ensure a fair comparison, we trained three available frameworks on our large BiGe dataset and compare the results. Specifically, we evaluate the Trimodal framework \cite{yoonSpeechGestureGeneration2020}, the SEEG framework \cite{liangSEEGSemanticEnergized2022}, and the HA2G framework \cite{liuLearningHierarchicalCrossModal2022a} against our proposed model. Regrettably, we were unable to compare our model to the model proposed by \citet{aoRhythmicGesticulatorRhythmAware2022a} as we did not get access to their implementation.

We adhered to the original configurations for each framework and only altered the output dimension to accommodate the higher joint dimensionality in our dataset. We measured the performance of each model using the Fréchet Gesture distance (FGD) \cite{yoonSpeechGestureGeneration2020}, the Diversity distance \cite{liuLearningHierarchicalCrossModal2022a}, and the mean of the absolute errors between the generated joint positions and ground truth (MAJE). As the Diversity distance exhibited high variance, we report the average of one thousand calculations. The results are reported in Table~\ref{tbl:res}.

The experimental results demonstrate that our model surpasses the existing frameworks in all evaluated metrics. Notably, the AQ-GT framework achieved a three times lower FGD than the HA2G. For the Diversity distance, our framework surpasses the distance of all other frameworks and even exceeds the human ground truth. It is important to note, that the diversity measures the absolute distance between random pairs of gesture samples based on their latent representations. Our evaluation result, while encouraging, may thus imply the possibility that the discrete representation of gestures could introduce outliers that are mapped in different regions of the latent space, therefore increasing the Diversity distance. 

To compare our model to a wider range of published frameworks, we additionally trained the model on the TED gesture dataset \cite{yoonRobotsLearnSocial2019a}. Our model architecture and configuration were kept unchanged, and only the input and output dimensions were adjusted to accommodate the differing body joint dimensions. To prevent an unfair advantage, we also pre-trained a new $VQ_G$ model on the TED dataset and incorporated it during the training process of our framework. The results are given in Table~\ref{tbl:ted_res}, again illustrating a significant improvement in FGD compared to all other approaches. Additionally, our framework achieves the second-highest Diversity distance score, closely trailing the HA2G framework. 

\subsection{Human Rater Study}
As assessing the quality of co-speech gestures is highly subjective, gesture generation frameworks are commonly evaluated, by employing human raters  \cite{nyatsangaComprehensiveReviewDataDriven2023b}. We therefore conducted a human rater study to evaluate our proposed framework in comparison to the frameworks by \citet{liuLearningHierarchicalCrossModal2022a} and \citet{liangSEEGSemanticEnergized2022}. To this end, we randomly selected ten 20-second-long audio/text sequences from our test set and generated co-speech gestures using all three models. The study was conducted as an online evaluation with 70 English-speaking participants (50\% female and 50\% male). Participants had to watch the videos in random order and, using a 5-point Likert scale, rate the co-speech gestures concerning human likeness and smoothness of motion as well as how much it matched the accompanied speech in content and timing (beat). Prior to the main study, we conducted a pre-survey with 10 participants to estimate the time required for each sequence. Based on the results of this pre-study, we excluded in the analysis any participant who remained less than 30 seconds on the survey pages, as well as any participant that deviated more than two times the standard deviation from the mean time. After applying these conditions, 39 participants remained. As our results did not fulfill the normality and homogeneity of variances tests, we used a generalized linear mixed-effects model (GLMM) for a repeated measure analysis. To compare the pairwise significance effect of the different models, we then used the Dunn posthoc test. The results are presented in Figure~\ref{fig:study_result},  demonstrating that the AQ-GT approach proposed here significantly outperforms both the HA2G and SEEG frameworks in all rated categories. Notably, our framework achieves ratings of human likeness and smoothness comparable to human ground truth (no significant difference), while the ratings of content and beat matching are still significantly lower than ground truth. These findings suggest that while our approach is capable of generating appropriate, smooth, and natural gestures, it still falls short in terms of generating convincing gestures that accurately reflect the content of speech or convey substantial additional meaning.

\begin{table}[t]
  \centering
  \caption{Comparison of the proposed original model with four different versions where single components were left out (for FGD and MAJE lower is better; for Diversity higher is better).}
  \begin{tabular}{lcccc}
    \toprule
    Methods & FGD $\downarrow$ & Diversity $\uparrow$ & MAJE $\downarrow$\\\
    Ground Truth & 0.00 & 42.128 & 0.00\\
    \midrule
    \textbf{w/o GRU} & 57.75 & 16.235 & 0.0330\\
    \textbf{w/o Temporal Aligner} & 1.653 & 39.054 & 0.0157\\
    \textbf{w/o VQ\textsubscript{G}} & 1.342 & 41.698 & 0.0102\\
    
    \textbf{w/o Transformer} & 0.777 & 43.592 & 0.0086\\
    \midrule
    \textbf{Original} & \textbf{0.3977} & \textbf{42.885} & \textbf{0.0078}\\ 
    \bottomrule
  \end{tabular}
  \label{tbl:ablation}
\end{table}

\section{Ablation Study}
In order to assess the contribution of individual components of our proposed framework, we conducted an ablation study employing four different architectural configurations. The impact of the GRU-Transformer was evaluated by selectively removing all GRU layers (w/o GRU) or all Transformer layers (w/o Transformer). The Temporal Aligner was evaluated by removing either the final reconstruction of gestures with the $VQ_G$ (w/o VQ\textsubscript{G}) or the entire Temporal Aligner (w/o Temporal Aligner). Each modified architecture was re-trained for 50 epochs, and the best-performing validation loss weights were restored. The performance on the test set is reported in Table~\ref{tbl:ablation}. As can be seen, the removal of the GRU layers had the largest impact on all reported measures. Visual inspection of the generated output did also show strongly deformed joints without any visible gestures. On the other hand, the removal of the Transformer layer did lead to a small degradation in the FGD and MAJE but led to a higher Diversity distance value. As argued above, this could indicate problems with the Diversity distance, as it is highly susceptible to noise and outliers in the gestures. Nevertheless, this shows that the Transformer layer did have the intended effect of learning long-term dependencies between frames and reducing noise in the generated gestures. 

The removal of the $VQ_G$ reconstruction led to a strong degradation in all measures but still achieved results similar to the HA2G framework (see Table~\ref{tbl:res}). Removing the entire Temporal Aligner leads to an even more substantial decrease in performance, achieving results more similar to the Trimodal and SEEG frameworks. This clearly indicates that learning the temporal dependencies of the resulting intermediate sequence helps to create more coherent and natural gestures and generally should be incorporated in future co-speech gesture synthesis frameworks.

\section{Conclusion}
In this paper, we proposed a novel co-speech gesture synthesis framework, called  AQ-GT, which combines and utilizes several approaches and techniques for the generation of co-speech gestural motion. Most notably, we combine GRU and transformer layers to model the non-linear temporal dependencies in multimodal behavior. In addition, we use a vector quantized autoencoder to allow for a discrete latent space representation for sequential data in both modalities (speech and gesture), and employ a dedicated Temporal Aligner network to reconstruct natural gestures. For training this complex model, we collected a large set of highly accurate multimodal data, using an extended data processing pipeline to track full 3D body joints from monocular videos. Our evaluations study results show that this approach is able to synthesize co-speech gestures that surpass the current state-of-the-art both in objective and subjective measures, and are in parts indistinguishable from human gesturing in terms of human-likeness and smoothness. By performing the ablation study, we were able to 
show that every component has a beneficial influence on the objective measures.

Overall, the proposed AQ-GT framework contributes to the field and advances the current state-of-the-art in co-speech gesture synthesis. Although our framework is able to create convincing natural gestures, there are still challenges. For one, the Diversity distance showed that there are possibly artifacts in the gestures generated from the discrete latent representation. In addition, the subjective study revealed that the present framework (just as all other learning-based approaches) is still unable to cover the meaning dimension of human communication and therefore generates primarily conversational "motor gestures" that support the act of speaking but rarely provide additional information. Another drawback that is inherent in learning-based co-speech gesture synthesis, is the inability to guide or control the generation of gestures. In future work we will both look at the drawbacks of discrete representations and how these can be improved, as well as if it is possible to guide the generation of gestures to more closely match the content of the conversation by conveying additional meaning. In current work, published elsewhere, we have started to extend the AQ-GT model to support form and meaning target features in gesture synthesis, learned from a richly annotated but smaller dataset. This shows that the proposed AQ-GT model can provide a suitable basis for further explorations into the generation of not just co-speech motion, but really meaningful gestures as an integral part of multimodal communicative behavior.

\clearpage
\bibliographystyle{ACM-Reference-Format}
\bibliography{main}

\end{document}